\documentclass[12pt]{article}
\usepackage{amsmath}
\usepackage{amssymb}
\usepackage{amsthm}
\usepackage[textwidth=17cm,textheight=24cm]{geometry}
\newcommand{\N}{\ensuremath{\mathbb{N}}}
\newcommand{\C}{\ensuremath{\mathbb{C}}}

\newcommand{\sdet}{\ensuremath{\operatorname{sdet}}}
\newcommand{\res}{\ensuremath{\operatorname{res}}}

\newcommand{\bt}{\ensuremath{{\boldsymbol{t}}}}

\newcommand{\btau}{\ensuremath{{\boldsymbol{\tau}}}}
\newtheorem{The}{Theorem}
\newtheorem{Cor}{Corolary}

\newtheorem{Pro}{Proposition}
\newtheorem{Lem}{Lemma}

\begin{document}

\title{Darboux Transformations for\\ Super-Symmetric KP Hierarchies}

\author{Q.P. Liu$^1$ and Manuel Ma\~{n}as$^2$\\[.3cm]
$^1$Beijing Graduate School,
China University of Mining and Technology\\
Beijing 100083, China\\[0.2cm]
$^1$The Abdus Salam International Centre for Theoretical Physics\\
34100 Trieste, Italy\\[.2cm] $^2$Departamento de F\'\i sica
Te\'orica II,
Universidad Complutense\\
28040 Madrid, Spain
}

\date{}
\maketitle
\begin{abstract}
We construct Darboux transformations for the super-symmetric KP
hierarchies of Manin--Radul and Jacobian types. 
We also consider the binary Darboux transformation for the
hierarchies. The iterations of both type of Darboux
transformations are briefly discussed.
\end{abstract}

\section{Introduction}
The first super-symmetric KP (SKP) hierarchy was introduced by
Manin and Radul \cite{manin} more than a decade ago. Another
relevant super-symmetrization is due to Mulase and Rabin
\cite{jacobi}. These systems have been  subject of extensive
studies from both mathematical and physical viewpoints.
 In particular we stress the possible applications in
two dimensional super-symmetric quantum gravity \cite{gao}. There
are now many results for SKP hierarchies such as the description
of their solution space in the framework of the universal
super-Grassmann manifold \cite{ueno}, the construction of their
additional symmetries \cite{addi} and tau functions \cite{luis}.

In our previous papers \cite{liu,lm}, we constructed the Darboux
transformations for the Manin--Radul's super-symmetric KdV (SKdV)
hierarchy and its reductions. We found that, as in ordinary case,
Darboux transformations constitute a very efficient tool for the
generation of solutions. The SKdV hierarchies are reductions of
SKP hierarchies, so a natural task is to construct Darboux
transformations for SKP hierarchies itself. It should be remarked
that such generalization is not so straightforward as one may
suppose. The reason is that the SKP hierarchies incorporates both
even time and odd time flows while the SKdV hierarchy only has the
even time flows. Very recently, Araytn {\em et al } have pointed
out that the natural candidate for the Darboux transformation does
not preserve the odd flows \cite{aratyn}.


The purpose of the paper is to present  proper Darboux
transformations for the SKP hierarchies. For that aim we must
recall, following Ueno and Yamada \cite{ueno}, that there two SKP
hierarchies, say SKP$_k$, $k=0,1$. In fact, by reversing the signs
of the odd times we go from $\text{SKP}_0$ to $\text{SKP}_1$. We
will show that the natural candidate for elementary Darboux
transformation is a map
$\text{SKP}_k\rightarrow\text{SKP}_{k+1},\,\mod 2$; thus, when
composed with a reversion of odd times gives new solution of the
$\text{SKP}_k$. Observe also that when this elementary Darboux
transformation is composed twice we get a transformation
$\text{SKP}_k\rightarrow\text{SKP}_k$. We also consider the binary
type of Darboux transformation for these hierarchies.

The paper is organized as follows. In the next section, we recall
all the relevant facts and formulae for the Manin--Radul and
Jacobian  SKP hierarchies. In \S 3 the reader may found our main
results. We construct both the so called elementary and binary
Darboux transformations. We show that while the binary one has a
rather straightforward generalization, the elementary one needs to
be modified. We discuss the iterations of both type of Darboux
transformations briefly.  Two appendices are included to present
some details.

\section{Super-symmetric KP Hierarchies}

In this section we remind the reader the basic aspects of the two
main SKP hierarchies: the Manin--Radul and Jacobian hierarchies.

The proper setting for the SKP hierarchies is the Sato's
formalism which we now recall  \cite{ueno,jacobi,luis}. We
consider the algebra $\mathcal{X}$ of super pseudo-differential
operators in the following form
\[
\mathcal{P}=\sum_{n\leq N}a_n(x,\theta, \bt, \boldsymbol{\tau})
D^n, \quad N\in\N
\]
here the coefficients $a_n$ are taken from a super-commutative
algebra
\[
\mathcal{Y}=\C[\!\![x, \bt]\!\!]\otimes \Lambda
(\theta,\boldsymbol{\tau})\otimes \mathcal{A}
\]
where $x$ is  an even variable, $\theta $ an odd variable, $\bt=\{
t_n\}_{n=1}^{\infty}$, $\boldsymbol{\tau}=\{
\tau_n\}_{n=1}^{\infty}$, and $\mathcal{A}$ is a finite or
infinite dimensional Grassmann algebra  over $\C$. The
super-differential operator $D$ is given by
\[
D=\theta\frac{\partial}{\partial
x}+\frac{\partial}{\partial\theta}
\]
which satisfies
\[
D^2=\partial:= \frac{\partial}{\partial x}, \quad D^{-1}=\theta
+\frac{\partial}{\partial \theta}\Big(\frac{\partial}{\partial
x}\Big)^{-1}.
\]

For any operator $L=\sum_i a_iD^i\in \mathcal{X}$ we denote the
   truncations to various orders  by
\[
L_+=\sum_{i\ge 0}a_iD^i,\quad L_-=\sum_{i<0}a_iD^i,\quad L_0=a_0
\]
and its super-residue is
\[
\res(L)=a_{-1}.
\]
We also follow the convention:
\begin{itemize}
\item Parentheses is intended to indicate that
an operator has acted on an argument.
\item Juxtaposition of
operators to indicate an operator product.
\end{itemize}

The parity of a super quantity $s$ is denoted by $|s|$; i.e., if
the quantity is  even, $|s|=0$, otherwise $|s|=1$.

To work out the so called binary Darboux transformation, we  need
the adjoint operation $^*$, which is defined as
\[
D^*=-D,\quad (D^n)^{\ast}=(-1)^{\frac{n(n-1)}{2}}D^n,\;\;\;
(fg)^*=(-1)^{|f||g|}g^*f^*
\]
for any $f, g\in \mathcal{X}$.

The super-vector fields on $\mathcal{Y}$ generating the SKP flows
are given by
\[
\partial_n:=\frac{\partial}{\partial t_n}, \quad
D_n:=\begin{cases}\dfrac{\partial}{\partial \tau_n}-\sum_{m\geq
1}\tau_m\dfrac{\partial}{\partial t_{n+m-1}}& \text{for the
Manin--Radul SKP,}\\ \dfrac{\partial}{\partial \tau_n}& \text{for
the Jacobian SKP.}
\end{cases}
\]
The action of $\partial_n$ on an object $f$ is sometimes written
as $f_{t_n}$.
They generate a  Lie super-algebra with relations
\begin{equation}
[\partial_n, \partial_m]=0,\quad [\partial_n, D_m]=0,\quad \{D_n,
D_m\}=\begin{cases}-2\partial_{n+m-1}& \text{for the Manin--Radul
SKP,}\\ 0& \text{for the Jacobian SKP,}\end{cases}
\end{equation}
where $[\cdot,\cdot]$ denotes the usual commutator and
$\{\cdot,\cdot\}$ the anti-commutator.

Now we introduce Sato's operator
\begin{equation}\label{W}
W=1+\sum_{n=1}^{\infty}w_n(x,\theta, \bt, \boldsymbol{\tau})
D^{-n},
\end{equation}
where the parity of the field $w_n $ is given by $(1-(-1)^n)/2 $;
i. e., $w_{2k}$ are even and $w_{2k-1}$ are odd, so that $W$ is a
homogeneous even operator.

Sato's equation for the SKP$_k$ hierarchies of type $k=0,1$ are
given by
\[
\begin{aligned}
\partial_n W&=-(W\partial^n W^{-1})_{-}W,\\
D_n W&=-(-1)^k(W A_n W^{-1})_{-}W,\quad \text{ with }
A_n:=\begin{cases}D^{2n-1}, \quad n\geq 1& \text{for the
Manin--Radul SKP,}\\ \dfrac{\partial}{\partial\theta}D^{2n-2},
\quad n\geq 1& \text{for the Jacobian SKP.}
\end{cases}
\end{aligned}
\]
Two observations are in order now: Notice first that given a Sato
operator $W(x,\theta,\bt,\btau)$ for SKP$_k$ the operator
$W(x,\theta,\bt,-\btau)$ is a Sato operator of SKP$_{k+1}$ (mod
2); i. e. both are essentially the same hierarchy. Secondly, for
the Jacobian hierarchy we have $A_n=D^{2n-1}-\theta D^{2n}$.

An essential ingredient in our forthcoming considerations are the
wave functions, which are solutions
$\varphi\in\mathcal{Y}\rightarrow\mathcal{A}$ of the following
equations
\begin{equation}
\partial_n\varphi =P_n\varphi,\quad
(-1)^kD_n\varphi=Q_n\varphi, \label{linear}
\end{equation}
where $P_n$ and $Q_n$ are defined in terms of the dressing
operator $W$ as follows
\[
P_n:=(W\partial^n W^{-1})_+,\quad Q_n:=(W A_n W^{-1})_+,
\]
 and accordingly we
also have their adjoint counterparts
\begin{equation}
\partial_n\varphi^*:=-P_{n}^{\ast}\varphi^*,\quad
(-1)^kD_n\varphi^*:=-Q_{n}^{\ast}\varphi^*. \label{adlinear}
\end{equation}

A second equivalent formulation of the SKP hierarchies is the
Zakharov--Shabat formalism \cite{ueno}, which can be regarded as
the compatibility condition of the linear systems \eqref{linear}
\begin{gather*}
\partial_n P_m-\partial_m P_n+[P_m,P_n]=0\\
\partial_n Q_m-(-1)^kD_mP_n+[Q_m,P_n]=0\\
(-1)^kD_mQ_n+(-1)^kD_nQ_m-\{Q_n,Q_m\}=\begin{cases}-2P_{n+m-1}&
\text{for the Manin--Radul SKP,}\\ 0& \text{for the Jacobian SKP.}
\end{cases}
\end{gather*}

For the Manin--Radul SKP hierachy there is a third formulation:
the Lax form. It  is obtained by dressing the operator $D$ with
$W$, getting thus the Lax operator
\[
L=WDW^{-1}=D+\sum_{n=1}^{\infty}u_n D^{-n+1}
\]
and the corresponding Lax representation
\begin{align*}
\partial_n L&=-[(L^{2n})_-,L]=[(L^{2n})_+, L], \\
(-1)^kD_n L&=-\{(L^{2n-1})_-, L\}=\{(L^{2n-1})_+, L\}-2L^{2n}.
\end{align*}

\section{Darboux Transformations}
In this section, we discuss the Darboux transformations for the
SKP hierarchy. We first consider so called elementary Darboux
transformation. We will see  that the naive candidate does not
qualify as a proper one, but by examining the failure we  show
how to remedy it and obtain a meaningful Darboux transformation.
The iteration of such transformation leads us to super
`Wro\'{n}ski' determinant representation for the solution of the
susy KP hierarchies. This is the generalization of the one found
by Ueno \emph{et al} \cite{ueno} in the framework of the
universal super-Grassmann manifold.

Next we discuss the binary Darboux transformation. We show that unlike the elementary one, the binary Darboux transformation does enjoy a straightforward generalization. As in the SKdV case, the iteration of such transformation provides us the solution of
 the SKP hierarchy in the form of super-Gramm determinants.
\subsection{Elementary Darboux Transformations}

We first introduce  odd \emph{gauge} operator
\begin{equation}
T:=\varphi D\varphi^{-1}=D-\varphi^{-1}(D\varphi) \label{T}
\end{equation}
where $\varphi$ is an even invertible wave function of the linear
system (\ref{linear}). Indeed, as shown in previous papers
\cite{liu,lm}, this very operator supplies us the Darboux
transformation for the  SKdV equation. We first observe
\begin{Lem}
The gauge operator  $T$ satisfies
\begin{gather*}
 (\partial_nT)T^{-1}=-(TP_nT^{-1})_-,\\
 (-1)^k(D_nT)T^{-1}=(TQ_nT^{-1})_-.
 \end{gather*}
\end{Lem}
\begin{proof}
First observe
\[
\partial_nT=-\partial_n(\varphi^{-1}(D\varphi))=-D(\varphi^{-1}\partial_n\varphi)
=-D(\varphi^{-1}P_n\varphi)
\]
and
\begin{align*}
(TP_nT^{-1})_-=&(\varphi D\varphi^{-1}P_n\varphi D^{-1}\varphi^{-1})_-\\
=&\varphi (D\varphi^{-1}P_n\varphi D^{-1})_-\varphi^{-1}\\
=&\varphi(D\varphi^{-1}P_n\varphi )D^{-1}\varphi^{-1}.
\end{align*}
Second,
\[
D_nT=-D_n(\varphi^{-1}(D\varphi))=D(\varphi^{-1}(D_n\varphi))
\]
and
\begin{align*}
(TQ_nT^{-1})_-=&(\varphi D\varphi^{-1}Q_n\varphi
D^{-1}\varphi^{-1})_-=
\varphi (D\varphi^{-1}Q_n\varphi D^{-1})_-\varphi^{-1}\\
=&\varphi (D\varphi^{-1}Q_n\varphi)D^{-1}\varphi.
\end{align*}
\end{proof}

This gauge operator allows us to define, in the realm of the
standard Darboux transformation, the transformed Sato's operator
\begin{equation}
\hat{W}:=TWD^{-1}.
\end{equation}
The susy KP flows on this transformed operator are
\begin{Lem}
The operator $\hat W$ satisfies
\begin{gather*}
\partial_n\hat{W}= -(\hat{W}\partial^n\hat{W}^{-1})_-\hat{W},\\
(-1)^{k+1}D_n\hat{W}=- (\hat{W}A_n\hat{W}^{-1})_-\hat{W}.
\end{gather*}
\end{Lem}
\begin{proof}
For  the even  evolution we have
\begin{align*}
\partial_n\hat{W}
=&-(TP_nT^{-1})_-TWD^{-1}-
T(W\partial^nW^{-1})_-WD^{-1}\\
=&(TP_nT^{-1})_+TWD^{-1}-TW\partial^nD^{-1}\\
=&(T(W\partial^nW^{-1})_+T^{-1})_+TWD^{-1}-TW\partial^nD^{-1}\\
=&(TW\partial^n W^{-1}T^{-1})_+TWD^{-1}-TW\partial^nD^{-1}\\
=& -(\hat{W}\partial^n\hat{W}^{-1})_-\hat{W},
\end{align*}
while for the odd time evolution
\begin{align*}
(-1)^kD_n\hat{W}
=&(TQ_nT^{-1})_-TWD^{-1}+
T(WA_nW^{-1})_-WD^{-1}\\
=&-(TQ_nT^{-1})_+TWD^{-1}+TWA_nD^{-1}\\
=&-(T(WA_nW^{-1})_+T^{-1})_+TWD^{-1}+TWA_nD^{-1}\\
=&-(TWA_n W^{-1}T^{-1})_+TWD^{-1}+TWA_nD^{-1}\\
=& (\hat{W}A_n\hat{W}^{-1})_-\hat{W}.
\end{align*}
\end{proof}
We see then that while the even evolutions are preserved by the
suggested transformation  the odd time evolutions are not.

At the level of wave functions for  any given solution $\psi$ of
\eqref{linear}  a new function $\hat\psi$ as
\begin{equation}
\hat{\psi}:=T\psi,\quad \hat{P}_n:=(TP_nT^{-1})_+,\quad
\hat{Q}_n=(TQ_nT^{-1})_+\label{Tr}
\end{equation}
then
\begin{Pro}
 The new function $\hat{\psi}$  satisfies
\begin{gather*}
\partial_n\hat{\psi}=\hat P_n\hat{\psi},\\
(-1)^{k+1}D_n\hat\psi=\hat Q_n\hat\psi.
\end{gather*}
\end{Pro}
\begin{proof}
We first compute
\[
\partial_n\hat{\psi}=((\partial_nT)T^{-1}+TP_nT^{-1})\hat{\psi}=(TP_nT^{-1})_+\hat{\psi}
\]
and then we calculate
\[
(-1)^kD_n\hat{\psi}=((-1)^k(D_nT)T^{-1}-TQ_nT^{-1})\hat{\psi}
=-(TQ_nT^{-1})_+\hat{\psi}.
\]
\end{proof}

Hence, we observe that is true that the SKP$_k$ odd flows will not
be preserved because there is a change $\tau_n\to-\tau_n$ involved
within the transformation. In fact, this Darboux transformation
generates a map $\text{SKP}_{k}\rightarrow \text{SKP}_{k+1}$ from
solutions of the $\text{SKP}_{k}$ to solutions of
$\text{SKP}_{k+1}$, $k\mod 2$. That is:
\begin{The}
Given a Sato operator $W$ of the $\text{SKP}_{k}$ hierarchy then
$\hat W$ is a Sato operator of the $\text{SKP}_{k+1}$ ($k\, \mod 2
$).
\end{The}

Moreover, for
\[
\tilde W(x,\theta,\bt,\btau):=\hat W(x,\theta,\bt,-\btau)
\]
we have
\begin{Cor}
Given a Sato operator $W$ of the $\text{SKP}_{k}$ hierarchy then
$\tilde W$ is again a Sato operator of the $\text{SKP}_{k}$
hierarchy.
\end{Cor}

Let us remark that, on the level of Lax formulation of the
Manin--Radul SKP$_0$ hierarchy, this point is treated in \S III
of \cite{aratyn}. Their observation is that the Darboux
transformation does  preserve only the bosonic flows. But, as we
have seen this can be remedied  by reversing the sign of the
fermionic flows. However, there is even another solution to this
problem: the idea is that the single step of the transformation
preserves the even flows but changes the odd flows in the way
$\tau_n$ $\to$ $-\tau_n$. Therefore, we do successively two steps
of such transformation which do preserve the odd flows as well.

Given two distinct solutions of the system \eqref{linear}
$\theta_0$ (even) and $\theta_1$ (odd), we construct an operator
\begin{equation}
T_e=\partial +\alpha D+a
\end{equation}
where the coefficients $\alpha$ and $a$ are given in terms of $\theta_0$ and
$\theta_1$
\begin{equation}
\alpha= D\ln\sdet\mathcal{ F},\quad
a=-\frac{\sdet\hat\mathcal{F}}{\sdet\mathcal{F}} \label{aa}
\end{equation}
with the  super-matrices $\mathcal{F}$ and $\hat\mathcal{F}$
defined as
\[
\mathcal{F}:=\begin{pmatrix}
  \theta_0 & \theta_1 \\
  D\theta_0 & D\theta_1
\end{pmatrix},\quad
\Hat{\mathcal{F}}:=\begin{pmatrix}
  \partial\theta_0 & \partial\theta_1 \\
  D\theta_0 & D\theta_1
\end{pmatrix}.
\]
Here sdet means the super-determinant or Berezian of a
super-matrix.

We define as well
\begin{equation}\label{10}
\hat{W}:=T_eW\partial^{-1}
\end{equation}
so that
\begin{gather*}
 \hat{w}_1:=w_1+\alpha,\quad\hat{w}_2:=w_2-\alpha w_1+a,
\\[0.2cm] \hat{w}_{n+2}:=w_{n+2}+(-1)^{|w_{n+1}|}\alpha w_{n+1}+a
w_n+\alpha (Dw_n)+w_{nx}, \quad (n\geq 1).
\end{gather*}

Then
\begin{The}
The operator $\hat{W}$ defined by \eqref{10} satisfies the SKP$_k$
hierarchy
\[
\partial_n \hat{W}=-(\hat{W}\partial^n
\hat{W}^{-1})_{-}\hat{W},\quad D_n \hat{W}=-(\hat{W} A_n
\hat{W}^{-1})_{-}\hat{W}, \quad (n\geq 1).
\]
\end{The}
So in the SKP case, the proper elementary Darboux transformation
is generated by the compound operator $T_e$.

This Darboux transformation can be iterated and the solutions can
be formed in terms of super `Wro\'nski' determinants as in super
KdV case.
\begin{Pro}
Given 2n distinct solutions $\theta_i$ $ (i=0, \cdots, 2n-1)$ of the linear system (\ref{linear}) with parities $|\theta_i|=(-1)^i$, then the new Sato's operator
is given by
\begin{equation}
\hat{W}=T_{e}[n]W\partial^{-n},\;\;\; T_{e}[n]=\partial^n+\sum_{i=0}^{2n-1}a_iD^i
\label{te}
\end{equation}
where the coefficients $a_i$ of the operator $T_{e}[n]$ are given by solving the linear equations
\begin{equation}
T_{e}[n](\theta_i)=0, \quad i=0, \cdots, 2n-1. \label{algebra}
\end{equation}
\end{Pro}

To obtain the explicit transformations between fields, one has to
solve the linear algebraic system \eqref{algebra} first, then
compare the coefficients of the different powers $D^{i}$ of the
equation \eqref{te}. The system \eqref{algebra} can be solved
easily as we did in \cite{liu,lm}. To this end, we introduce new
variables
\begin{gather*}
\boldsymbol{a}^{(0)}:=(a_0, a_2, \cdots , a_{2n-2}),\quad
\boldsymbol{a}^{(1)}:=(a_1, a_3, \cdots , a_{2n-1}),\\
\boldsymbol{\theta}^{(0)}:=(\theta_0,\theta_2, \cdots ,
\theta_{2n-2}),\quad
\boldsymbol{\theta}^{(1)}:=(\theta_1,\theta_3, \cdots , \theta_{2n-1}),\\
\boldsymbol{b}^{(i)}:=\partial^n \boldsymbol{\theta}^{(i)},
\quad\mathcal{W}^{(i)}:=\begin{pmatrix}\boldsymbol{\theta}^{(i)}\\
\partial\boldsymbol{\theta}^{(i)}\\ \vdots \\ \partial^{n-1}\boldsymbol{\theta}^{(i)}
\end{pmatrix}, \quad i=0, 1
\end{gather*}
so that the linear algebraic system \eqref{algebra} can be
reformulated as
\begin{equation}
(\boldsymbol{a}^{(0)}, \boldsymbol{a}^{(1)}){\cal
W}=-(\boldsymbol{b}^{(0)},\boldsymbol{b}^{(1)})
\end{equation}
where
\[
\mathcal{W}=\begin{pmatrix} \mathcal{W}^{(0)}&\mathcal{W}^{(1)}\\
D\mathcal{W}^{(0)}&D\mathcal{W}^{(1)}
\end{pmatrix}
\]
Then, we obtain
\begin{gather*}
a_{2i+2}=-\frac{\det\Big(\mathcal{W}^{(0)}_{i}-\mathcal{
W}^{(1)}_{i}
(D\mathcal{W}^{(1)})^{-1}
(D\mathcal{W}^{(0)})\Big)}{
\det\Big(\mathcal{W}^{(0)}-\mathcal{W}^{(1)}
(D\mathcal{W}^{(1)})^{-1}(D\mathcal{W}^{(0)})\Big)}=-\frac{\sdet\mathcal{W}_i}{
\sdet\mathcal{W}},\quad i=1, \cdots, n-1, \label{a0}\\
a_{2i-1}=-\frac{\det\Big((D\mathcal{W}^{(1)})_i-(D\mathcal{W}^{(0)})_i(\mathcal{W}^{(0)})^{-1}\mathcal{W}^{(1)}\Big)}
{\det\Big(
D\mathcal{W}^{(1)}-(D\mathcal{W}^{(0)})(\mathcal{W}^{(0)})^{-1}\mathcal{W}^{(1)}
\Big)}, \quad i=1, \cdots, n. \label{a00}
\end{gather*}
where the matrix $\mathcal{W}^{(j)}_{i}$ is the
$\mathcal{W}^{(j)}$ with its $i$-th row replaced by
$\boldsymbol{b}^{(j)}$ and $\mathcal{W}=\begin{pmatrix}
  \mathcal{W}_i^{(0)} & \mathcal{W}_i^{(1)} \\
  D\mathcal{W}^{(0)} & D\mathcal{W}^{(1)}
\end{pmatrix}$, the matrix $(D\mathcal{W}^{(1)})_i$ is the matrix $
D\mathcal{W}^{(1)}$ with its $i$-th row replaced by
$\boldsymbol{b}^{(1)}$ and the matrix $(D\mathcal{W}^{(0)})_i$ is
the matrix $ D\mathcal{W}^{(0)}$ with its $i$-th row replaced by
$\boldsymbol{b}^{(0)}$.

By considering the equation \eqref{te}, we obtain the general
formulae
\begin{gather*}
\hat w_{2n-k}=\sum^{2n}_{j=k}a_j\sum_{i=0}^{j-k}C_{i,j-k-i,j},
\quad
k=1, \cdots, 2n,\\
\hat w_{2n+k}=\sum^{2n}_{j=0}a_j\sum_{i=0}^{j}C_{i,j+k-i,j},\quad
k=0, 1, \cdots
\end{gather*}
with $w_0=a_{2n}=1$ and the $a_i$ are given by
\eqref{a0}-\eqref{a00} and
\[
C_{i,j,k}=\begin{bmatrix}k\\ k-i
\end{bmatrix}(-1)^{|w_j|(k-i)} (D^iw_j)
\]
where $\begin{bmatrix}\cdot \\ \cdot\end{bmatrix}$ denotes the
super-binomial coefficients.

\subsection{Binary Darboux Transformation}
In this section we consider the extension of the well known binary
Darboux transformation for the KP hierarchy to its
super-symmetrizations.

For an eigenfunction $\varphi$ and an adjoint  eigenfunction
$\varphi^*$ satisfying the linear system \eqref{linear} and
\eqref{adlinear} respectively, we may introduce a potential
operator $\Omega$ as follows
\begin{equation}\label{omega}
\begin{gathered} D\Omega(\varphi^*,\varphi)= \varphi^*\varphi,
\partial_n\Omega=\res(D^{-1}\varphi^*  P_n\varphi D^{-1}),\quad
D_n \Omega=\res(D^{-1}\varphi^* Q_n\varphi D^{-1})
\end{gathered}
\end{equation}
where we choose $\varphi$ as an even quantity and $\varphi^*$ as
an odd one, so that our potential operator $\Omega$ is even.

By lengthy calculation, see Appendix A, one can show that $\Omega$
is well defined,
 i.e., the equations \eqref{omega}  are consistent.

In terms of  $\Omega$, we introduce the \emph{gauge} operator
\begin{equation}
T:=1-\varphi\Omega^{-1}D^{-1}\varphi^* \label{biT}
\end{equation}
where we assume the invertibility of $\Omega$. The formal inverse of $T$ is
\begin{equation}
T^{-1}=1+\varphi D^{-1}\Omega^{-1}\varphi^*.
\end{equation}
In Appendix B we prove that
 \begin{Pro}\label{appendixb}
 The gauge operator $T$ solves
\begin{equation}
(\partial_nT)T^{-1}=-(TP_nT^{-1})_-,\quad
(D_nT)T^{-1}=-(TQ_nT^{-1})_-. \label{gg}
\end{equation}
  \end{Pro}
  This property qualifies $T$ for generating a Darboux transformation
 \cite{oevel}.
Given a Sato operator $W$ we introduce its transformed $\hat W$ as
follows
\begin{equation}\label{bi}
\hat{W}:=TW.
\end{equation}
\begin{Pro}
The operator $\hat W$ is again a Sato operator.
\end{Pro}
\begin{proof}
We begun with
\begin{align*}
\partial_n\hat{W}=&
=-(TP_nT^{-1})_-TW-T(W\partial^n W^{-1})_-W\\ = &
(TP_nT^{-1})_+TW-TP_nW+T(W\partial^n W^{-1})_+W-TW\partial^n\\ = &
(T(W\partial_nW^{-1})_+T^{-1})_+TW-TW\partial^n\\ = &
(TW\partial^n W^{-1}T^{-1})_+TW-TW\partial^n\\ = &
-(\hat{W}\partial^n\hat{W}^{-1})_-\hat{W}
\end{align*}
and
\begin{align*}
D_n\hat{W} &=
-(TQ_nT^{-1})_-TW-T(WA_nW^{-1})_-W\\
&=(TQ_nT^{-1})_+TW-TQ_nW+T(WA_nW^{-1})_+W-TWA_n\\
&=(T(WA_nW^{-1})_+W^{-1})_+TW-TWA_n\\
&=-(\hat{W}A_n\hat{W}^{-1})_-\hat{W}
\end{align*}
Therefore, the transformed Sato's operator indeed is the solution
of Sato's equation.
\end{proof}

The explicit transformation is
\begin{gather*}
\hat{w}_1=w_1+\varphi\Omega^{-1}\varphi^*,\\
 \hat{w}_2=w_2-\varphi\Omega^{-1}\Big((D\varphi^*) +\varphi^* w_1\Big),\\
\hat{w}_{2j+1}=w_{2j+1}+(-1)^{j}\varphi\Omega^{-1}\Bigl({\varphi^*}^{(j)}+
\sum_{k=0}^{j-1}(-1)^{k}\Bigl((D\varphi^*
w_{2k+1})^{(j-k-1)}-(\varphi^* w_{2k+2})^{(j-k-1)}\Bigr),\\
\begin{split}
\hat{w}_{2j+2}=w_{2j+2}+(-1)^{j+1}\varphi\Omega^{-1}\Big((D\varphi^*)^{(j)}+
(\varphi^* w_1)^{(j)}+ \sum^{j}_{k=1}(-1)^k\Bigl((D\varphi^*
w_{2k})^{(j-k)}\\ +(\varphi^* w_{2k+1})^{(j-k)}\Bigr)\Big),
\end{split}
\end{gather*}
for $j=1, 2, \cdots$ and where
$f^{(i)}=\dfrac{\partial^if}{\partial x^i}$.

For   any given solution $\psi$ of \eqref{linear} we define
\begin{equation*}
\hat{\psi}:=T\psi,\quad \hat{P}_n:=(TP_nT^{-1})_+,\quad
\hat{Q}_n=(TQ_nT^{-1})_+
\end{equation*}
and as before
\begin{Pro}
 The  function $\hat{\psi}$  satisfies
\begin{gather*}
\partial_n\hat{\psi}=\hat P_n\hat{\psi},\\
D_n\hat\psi=\hat Q_n\hat\psi.
\end{gather*}
\end{Pro}

For the Manin--Radul SKP hierarchy the transformation on the Lax
level can be easily obtained as
\[
\hat{L}=TLT^{-1}
\]
and we found that  $\hat{L}$ satisfies
\[
\hat{L}_{t_n}=[(\hat{L}^{2n})_+, \hat{L}],\quad
D_n\hat{L}=\{(\hat{L}^{2n-1})_+,\hat{L} \}-2\hat{L}^{2n}.
\]
Indeed, we have
\begin{align*}
\hat{L}_{t_n}=&[T_{t_n}T^{-1}+T(L^{2n})_+T^{-1},\hat{L}]\\
=&[-(TP_nT^{-1})_-+T(L^{2n})_+T^{-1}, \hat{L}]\\
=&[(TP_nT^{-1})_+, \hat{L}]\\
=&[(TL^{2n}T^{-1})_+,\hat{L}]=[(\hat{L}^{2n})_+,\hat{L}]
\end{align*}
since $(T(L^{2n})_-T^{-1})_+=0 $

Likewise,
\begin{align*}
D_n\hat{L}=&\{(D_nT)T^{-1}+T(L^{2n-1})_+T^{-1},
\hat{L}\}-2\hat{L}^{2n}\\
=&\{-(TQ_nT^{-1})_-+T(L^{2n-1})_+T^{-1},\hat{L}\}-2\hat{L}^{2n}\\
=&\{(T(L^{2n-1})_+T^{-1})_+,\hat{L}\}-2\hat{L}^{2n}\\
=&\{(\hat{L}^{2n-1})_+,\hat{L}\}-2\hat{L}^{2n}
\end{align*}
because of $(T(L^{2n-1})_-T^{-1})_+=0 $. The binary Darboux
transformation can be iterated. To do this, we need to consider
the effect of this transformation for adjoint wave functions
$\psi^*$. It is indeed not difficult to find they are transformed
as follows
\[
\hat{\psi}^*=\psi^*
-\varphi^*(\Omega(\varphi^*,\varphi))^{-1}\Omega(\psi^*,\varphi).
\]
Another important fact is the following
\[
\Omega(\hat{\psi^*},\hat{\psi})=\Omega(\psi^*,\psi)-\Omega(\psi^*,\varphi)
\Omega(\varphi^*,\varphi)^{-1}\Omega(\varphi^*,\psi)
\]
then as in the pure bosonic case \cite{os}, by iterations of the binary
Darboux transformation leads to the following result
\begin{Pro}
Let $\varphi_i$ $(i=1, 2,\cdots, n)$ be n bosonic solutions of
(\ref{linear}) and $\varphi^*_i$ $(i=1, 2,\cdots, n)$ be n
fermionic solutions of (\ref{adlinear}), let
$\mathcal{G}=(\Omega(\varphi^*_i,\varphi_j))$be a Gramm type of
matrix and $\mathcal{G}_j$ be the matrix $\mathcal{G}$ with its
$i$-th row replaced by $(\varphi_1, \cdots, \varphi_n)$. Define
\[
T[n]=1-\sum_{i=1}^{n}b_iD^{-1}\varphi^*_i, \quad
b_i=\frac{\det\;\mathcal{G}_i}{\det\;\mathcal{G}}.
\]
Then the new Sato's operator is given by
\begin{equation}
\hat{W}=T[n]W
\label{itr}
\end{equation}
\end{Pro}
From (\ref{itr}) we obtain the explicit transformation on the level of fields
\begin{gather*}
\hat{w}_1=w_1+\sum_{j=1}^{n}b_j\varphi^*_j,\\
\hat{w}_2=w_2-\sum_{j=1}^{n}b_j\Big((D\varphi^*_j)+\varphi^*_j
w_1\Big),\\ \hat{w}_{2j+1}=w_{2j+1}+(-1)^{j}
\sum_{s=1}^{n}b_s\Big({\varphi^*_s}^{(j)}
+\sum_{k=0}^{j-1}(-1)^k\Big((D\varphi^*_sw_{2k+1})^{(j-k-1)}
-(\varphi^*_sw_{2k+2})^{(j-k-1)}\Big)\Big)\\
\begin{split}\hat{w}_{2j+2}=w_{2j+2}+(-1)^{j+1}
\sum_{s=1}^{n}b_s\Big((D\varphi^*_s)^{(j)}+ (\varphi^*_s
w_1)^{(j)}+ \sum_{k=1}^{j}(-1)^k\Big((D\varphi^*_sw_{2k})^{(j-k)}
\\+(\varphi^*_s w_{2k+1}^{(j-k)})\Big)\Big), \end{split}
\end{gather*}
for $j=1, 2, \cdots$

\section*{Acknowledgments}

QPL should like to thank the Abdus Salam International Centre for
the Theoretical Physics for hospitality and support. QPL  is
partially supported  by the National Natural Science Foundation of
China (grant no. 19971094). MM is partially supported by Comisi\'{o}n
Inter-ministerial de Ciencia y Tecnolog\'{\i}a PB98--0821.

\vspace{.6cm}

\begin{appendix}

\appendix{\textbf{Appendix A: Compatibility for the potential.}}
\setcounter{equation}{0}
\def\theequation{A.\arabic{equation}}
We first list some useful identities (see also \cite{shaw})
\begin{gather}
(\Lambda^*)_+=(\Lambda_+)^*,\quad (\Lambda^*)_-=(\Lambda_-)^*,
\label{a1}\\
(D^{-1}\Lambda)_-=D^{-1}(\Lambda^*)_0+D^{-1}(\Lambda_-),\quad
(\Lambda D^{-1})_-=\Lambda_0D^{-1}+\Lambda_-D^{-1}, \label{a2}\\
\res(\Lambda)=\res(\Lambda^*), \label{a3}\\ D(\res\Lambda)=
\res(D\Lambda-(-1)^{|\Lambda|}\Lambda D), \label{a4}\\ \res
(\Lambda D^{-1})=(\Lambda)_0, \quad \res
(D^{-1}\Lambda)=(-1)^{|\Lambda|}(\Lambda^*)_0, \label{a5}\\
\res(D^{-1}\Lambda_1\Lambda_2
D^{-1})=\res(D^{-1}(\Lambda_{1}^*)_0\Lambda_2
D^{-1})+\res(D^{-1}\Lambda_1(\Lambda_2)_0D^{-1}), \label{a6}
\end{gather}
where $\Lambda$ is an arbitrary super-pseudo-differential operator
and $\Lambda_i$ are assumed as arbitrary super-differential
operators. These identities can be verified easily.

First we show
\begin{equation}
\partial_m\partial_n\Omega=
\partial_n\partial_m\Omega.
\label{a8}
\end{equation}
Indeed,
\begin{align*}
\partial_m\partial_n\Omega=&
\res(D^{-1}\varphi^*_{t_m} P_n\varphi D^{-1})+\res(D^{-1}\varphi^*
P_{n,t_m}\varphi D^{-1})+\res(D^{-1}\varphi^* P_n\varphi_{t_m}
D^{-1})\\ =& -\res(D^{-1}(P_{m}^{*}\varphi^*)P_n\varphi D^{-1})+
   \res(D^{-1}\varphi^* P_{n,t_m}\varphi D^{-1})+
   \res(D^{-1}\varphi^* P_n(P_m\varphi) D^{-1}).
\end{align*}
Thus, we have
\begin{align*}
(\partial_m\partial_n-\partial_n\partial_m)\Omega=&
  \res(D^{-1} (P_{n}^{*}\varphi^*)P_m\varphi D^{-1})-
   \res(D^{-1}(P_{m}^{*}\varphi^*)P_n\varphi D^{-1})\\\begin{split}
      +
   \res(D^{-1}(\varphi^* P_{n,t_m}-P_{m,t_n})\varphi D^{-1})+
   \res(D^{-1}\varphi^* P_n(P_m\varphi) D^{-1})\\
   -\res(D^{-1}\varphi^* P_m(P_n\varphi)D^{-1})\end{split}\\
  =& \res(D^{-1}\varphi^*([P_n,P_m]+P_{n,t_m}-P_{m, t_n})\varphi D^{-1})=0
\end{align*}
where, in particular, we have used the identity (\ref{a6}) twice
with $\Lambda_1=\varphi^* P_n $ , $\Lambda_2=P_m\varphi$ and
$\Lambda_1=\varphi^* P_m$, $\Lambda_2=P_n\varphi$, respectively.
Therefore, (\ref{a8}) holds.

Next we prove
\begin{equation}
D_m\partial_n\Omega=\partial_nD_m\Omega. \label{a9}
\end{equation}

To this end, we calculate
\begin{align*}
D_m\partial_n\Omega=& -\res(D^{-1}(D_m\varphi^*)P_n\varphi D^{-1})
+\res(D^{-1}\varphi^*(D_m P_n)\varphi D^{-1})
+\res(D^{-1}\varphi^* P_n(D_m\varphi)D^{-1})\\ =&
\res(D^{-1}(Q_{m}^{*}\varphi^*)P_n\varphi D^{-1})
+\res(D^{-1}\varphi^*(D_m P_n)\varphi D^{-1})
+\res(D^{-1}\varphi^* P_n(Q_m\varphi)D^{-1})
\end{align*}
and
\begin{equation*}
\partial_n D_m\Omega =
 -\res(D^{-1}(P^*_n\varphi^*)
 Q_m\varphi D^{-1})+\res(D^{-1}\varphi^* Q_{m,t_n}\varphi D^{-1})
+\res(D^{-1}Q_m(P_n\varphi)D^{-1}).
\end{equation*}
Thus,
\begin{align*}
D_m\partial_n\Omega-\partial_nD_m\Omega=&
\res(D^{-1}(Q_{m}^{*}\varphi^*)P_n\varphi D^{-1})
+\res(D^{-1}(P_{n}^{*}\varphi^*)Q_m\varphi D^{-1})\\ &
+\res(D^{-1}\varphi^* P_n(Q_{m}\varphi) D^{-1})-
\res(D^{-1}\varphi^* Q_{m}(P_n\varphi) D^{-1})\\ &
+\res(D^{-1}\varphi^*(D_m P_n-Q_{m, t_n})\varphi D^{-1})\\ =&
\res(D^{-1}\varphi^*([P_n, Q_m]+(D_mP_n-Q_{m,t_n})\varphi
D^{-1})=0;
\end{align*}
i.e., \eqref{a9} holds. Again, we used the formula \eqref{a6} with
$\Lambda_1=\varphi^* P_n$, $\Lambda_2=Q_m\varphi $ and
$\Lambda_1=\varphi^* Q_m$, $\Lambda_2=P_n\varphi$, respectively.

Finally, since
\begin{align*}
D_mD_n\Omega =& -\res(D^{-1}(D_m\varphi^*)Q_n\varphi D^{-1})+
\res(D^{-1}\varphi^*(D_mQ_n)\varphi D^{-1})\\
&-\res(D^{-1}\varphi^* Q_n(D_m\varphi )D^{-1})\\ =&
\res(D^{-1}(Q_{m}^{*}\varphi^*)Q_n\varphi D^{-1})+
\res(D^{-1}\varphi^*(D_mQ_n)\varphi D^{-1})\\ &-
\res(D^{-1}\varphi^* Q_n(Q_m\varphi)D^{-1})
\end{align*}
we obtain
\begin{align*}
(D_mD_n+D_nD_m)\Omega=& \res(D^{-1}(Q^{*}_{m}\varphi^*)Q_n\varphi
D^{-1})- \res(D^{-1}\varphi^* Q_n(Q_m\varphi) D^{-1})\\ &+
\res(D^{-1}(Q_{n}^{*}\varphi^*) Q_m\varphi)D^{-1})-
\res(D^{-1}\varphi^* Q_{m}(Q_n\varphi) D^{-1})\\ &+
\res(D^{-1}\varphi^*(D_mQ_n)\varphi D^{-1})
+\res(D^{-1}\varphi^*((D_nQ_m)\varphi D^{-1})\\ =&
-\res(D^{-1}\varphi^* Q_mQ_n\varphi D^{-1})-\res(D^{-1}\varphi^*
Q_nQ_m\varphi D^{-1})\\ &+ \res(D^{-1}\varphi^*(D_m
Q_n+D_nQ_m)\varphi D^{-1}) +D_nQ_m)\varphi D^{-1})\\  =&
\res(D^{-1}\varphi^*(D _mQ_n+D_nQ_m-\{Q_m,Q_n\} )\varphi D^{-1})\\
=&\begin{cases}-2\partial_{n+m-1}\Omega& \text{for the
Manin--Radul SKP,}\\ 0& \text{for the Jacobian SKP.}
\end{cases}
\end{align*}
Thus we conclude that the $\Omega $ given by \eqref{omega} is well
defined.\\

\appendix\textbf{Appendix B: Proof of Proposition \ref{appendixb}.}
\setcounter{equation}{0}
\def\theequation{B.\arabic{equation}}
In the one hand we have
\begin{align*}
T_{t_n}T^{-1} 
=&-(P_n\varphi)\Omega^{-1}D^{-1}\varphi^*
+\varphi\Omega^{-1}\res(D^{-1}\varphi^* P_n \varphi
D^{-1})\Omega^{-1}D^{-1}\varphi^* +\varphi
\Omega^{-1}D^{-1}(P_{n}^{*}\varphi^*)\\
&-(P_n\varphi)\Omega^{-1}D^{-1}(D\Omega -\Omega
D)D^{-1}\Omega^{-1}\varphi^*+
\varphi\Omega^{-1}D^{-1}(P_{n}^{*}\varphi^*)\varphi
D^{-1}\Omega^{-1}\varphi^* \\
&+\varphi\Omega^{-1}\res(D^{-1}\varphi^* P_n\varphi
D^{-1})\Omega^{-1}D^{-1}(D\Omega -\Omega
D)D^{-1}\Omega^{-1}\varphi^* \\ =& \varphi
\Omega^{-1}D^{-1}(P_{n}^{*}\varphi^*)
-(P_n\varphi)D^{-1}\Omega^{-1}\varphi^* \\
&+\varphi\Omega^{-1}\res(D^{-1}\varphi^* P_n\varphi
D^{-1})D^{-1}\Omega^{-1} \varphi^*
+\varphi\Omega^{-1}D^{-1}(P_{n}^{*}\varphi^*)\varphi
D^{-1}\Omega^{-1}\varphi^*.
\end{align*}

While on the other hand
\begin{equation*}
-(TP_nT^{-1})_-=
 -(P_n\varphi)D^{-1}\Omega^{-1}\varphi^* +
\varphi\Omega^{-1}D^{-1}(P_{n}^{*}\varphi^*)+
\varphi\Omega^{-1}(D^{-1}\varphi^* P_n\varphi
D^{-1})_-\Omega^{-1}\varphi^*.
\end{equation*}
Thus, we have
\begin{multline*}
T_{t_n}T^{-1}+(TP_nT^{-1})_-=
 \varphi\Omega\Big(\res(D^{-1}\varphi^* P_n\varphi D^{-1})D^{-1}+
D^{-1}(P_{n}^{*}\varphi^*)\varphi D^{-1}\\ -(D^{-1}\varphi^*
P_n\varphi D^{-1})_-\Big)\Omega^{-1}\varphi^*= 0,
\end{multline*}
where the identities \eqref{a2} and \eqref{a5} have been used.

We now turn our attention to  the odd flows,
\begin{align*}
(D_nT)T^{-1}
=&-(Q_n\varphi)\Omega^{-1}D^{-1}\varphi^*
+\varphi\Omega^{-1}\res(D^{-1}\varphi^* Q_n\varphi
D^{-1})\Omega^{-1}D^{-1}\varphi^*
-\varphi\Omega^{-1}D^{-1}(Q_{n}^{*}\varphi^*)\\
&-(Q_n\varphi)\Omega^{-1}D^{-1}(D\Omega -\Omega
D)D^{-1}\Omega^{-1}\varphi^*\\
&+\varphi\Omega^{-1}\res(D^{-1}\varphi^* Q_n\varphi
D^{-1})\Omega^{-1}D^{-1}(D\Omega -\Omega D
)D^{-1}\Omega^{-1}\varphi^*\\ &-\varphi
\Omega^{-1}D^{-1}(Q_{n}^{*}\varphi^*)\varphi
D^{-1}\Omega^{-1}\varphi^*\\ =&
-\varphi\Omega^{-1}D^{-1}(Q_{n}^{*}\varphi^*)
-(Q_n\varphi)D^{-1}\Omega^{-1}\varphi^*\\
&+\varphi\Omega^{-1}\res(D^{-1}\varphi^* Q_n\varphi
D^{-1})D^{-1}\Omega^{-1}\varphi^*
-\varphi\Omega^{-1}D^{-1}(Q_{n}^{*}\varphi^*)\varphi
D^{-1}\Omega^{-1}\varphi^*
\end{align*}
and
\begin{equation*}
(TQ_nT^{-1})_-
=\varphi\Omega^{-1}D^{-1}(Q_{n}^{*}\varphi^*)+(Q_n\varphi
)D^{-1}\Omega^{-1}\varphi^* -\varphi\Omega^{-1}(D^{-1}\varphi^*
Q_n\varphi D^{-1})_-\Omega^{-1}\varphi^*.
\end{equation*}

Thus by using \eqref{a2} we have
\begin{multline*}
(D_nT)T^{-1}+(TQ_nT^{-1})_- =
\varphi\Omega^{-1}\Big(\res(D^{-1}\varphi^* Q_n\varphi
D^{-1})D^{-1} -D^{-1}(Q_{n}^{*}\varphi^*)\varphi D^{-1}\\
-(D^{-1}\varphi^* Q_n\varphi D^{-1})_-\Big)\Omega^{-1}\varphi^*=0.
\end{multline*}
\end{appendix}


\end{document}